\documentclass[11pt]{article}
\usepackage{jhep}
\usepackage[english]{babel}
\usepackage{geometry}
\usepackage{babel}
\usepackage{amsmath}
\usepackage{amssymb}
\usepackage{physics}
\usepackage{bbm}

\title{On Complexity of Holographic Flavors}
\author[]{Francisco Jose Garcia Abad, Manuela Kulaxizi and Andrei Parnachev}
\affiliation[]{School of Mathematics, Trinity College Dublin, Dublin 2, Dublin}
\abstract{Quantum complexity of a thermofield double state in a strongly coupled quantum field theory has been argued to be holographically related
to the action evaluated  on  the Wheeler-DeWitt patch.
The growth rate of quantum complexity  in systems dual to Einstein-Hilbert gravity saturates a bound which
follows from the Heisenberg uncertainty principle.
We consider corrections to the  growth rate in models with flavor degrees of freedom.
They are realized by adding  a small number of flavor branes to the system.
Holographically, such corrections come from  the DBI action of the flavor branes evaluated on the  Wheeler-DeWitt patch.
We relate corrections to the growth of quantum complexity to  corrections to the mass of the system, and
observe that the bound on the growth rate is never violated. }

\begin{document}

\maketitle

\flushbottom
\section{Introduction and summary}\label{Introduction}
Quantum  complexity $\mathcal{C}$ is a quantity defined for a quantum  system, where 
unitary operations, called gates, are applied to pairs of qubits\footnote{We will consider the case of two-gates, but one can easily generalize the discussion to  the k-gates.}. 
For a state $\ket{\psi}$  complexity is defined as the minimum number of such gates that have to be applied to a simple reference state to produce $\ket{\psi}$.
It has  been argued \cite{Lloyd} that  due to the Heisenberg uncertainty principle quantum complexity obeys a bound on its growth rate:
\begin{equation}\label{eq:bound}
\frac{d \mathcal{C}}{dt}\leq \frac{2\cal{M}}{\pi},
\end{equation}
where $\mathcal{M}$ is the mass of the system (Check references \cite{Carmi:2017jqz}-\cite{Brown:2015lvg} for some violations of this bound) 

Recently a holographic recipe has been proposed \cite{Brown:2015bva, Brown:2015lvg} to compute complexity for thermofield double states in strongly coupled quantum field theories. 
(For related work, including a few lecture notes, see \cite{Prudencio:2014wpa}-\cite{Yang:2016awy}.)
The proposal of  \cite{Brown:2015bva, Brown:2015lvg}, which we refer to  as Complexity-Action (CA) proposal, 
makes use of the holographic representation of the thermofield double state in a strongly coupled quantum field theory
in terms of the eternal asymptotically AdS black hole \cite{Maldacena:2001kr}.
 On this spacetime one can define the Wheeler-DeWitt patch, shown in Fig. \ref{fig:WdWparts}. 
 The patch is anchored at boundary times $t_L$ and $t_R$, and the  proposal of   \cite{Brown:2015bva,Brown:2015lvg}
equates the complexity of the thermofield dual state $\ket{\psi(t_L,t_r)}$ with the action evaluated on the Wheeler-DeWitt patch $S_{WdW} $:
\begin{equation}
\label{proposalC}
\mathcal{C}(\psi(t_L,t_R))=\frac{S_{WdW}}{\pi \hbar},
\end{equation}
It was also shown in \cite{Brown:2015bva,Brown:2015lvg} that for the Einstein-Hilbert action, AdS black holes saturate the bound \eqref{eq:bound}.

In this paper we add massless matter in the fundamental representation to ${\cal N}=4$ super Yang-Mills and compute the corresponding corrections to $d\mathcal{C}/dt$.
We achieve this by adding a small number of flavor branes to the stack of the D3 branes. 
At strong 't Hooft coupling, we need to study flavor branes propagating in asymptotically $AdS_5\times S^5$ background.
The action of D-branes is  just the DBI action, and thus the CA correspondence identifies the correction to  quantum complexity
with the DBI action evaluated on the Wheeler-DeWitt patch
\begin{equation}
\label{correctionC}
 \delta \mathcal{C}=\frac{S_{DBI,WdW}}{\pi \hbar},
\end{equation}
Note that the variational problem for the DBI action is well defined and there is no need to introduce boundary terms in  \eqref{correctionC}.
 We will see that $\delta \mathcal{C}$  can be written as a function of temperature times the 
contribution of the flavor degrees of freedom to the total mass of the system,  $\delta M$.
One may wonder whether  the growth rate of the total quantum complexity still obeys the inequality \eqref{eq:bound},
\begin{equation}\label{eq:corr}
\frac{d \mathcal{C}_{tot}}{d t}\equiv \frac{d\mathcal{C}}{dt}+\frac{d (\delta \mathcal{C})}{dt}\stackrel{?}{\leq}\frac{2 \mathcal{M}_{tot}}{\pi}=2\left(\frac{\cal{M}}{\pi}+\frac{\delta \cal{M}}{\pi}\right).
\end{equation}
We will show that the corrections have the form
\begin{equation}
\frac{d(\delta \mathcal{C})}{dt}=- K(x) \frac{\delta \cal{M}}{\pi}, \quad \quad \quad x=\pi L T,
\end{equation}
with $K(x)$ a monotonically increasing function. It is important to note that this correction is negative because of the overall minus sign that appears in front of the Lorentzian DBI action.
Hence, the flavor corrections  reduce the rate at which complexity grows and the bound  \eqref{eq:bound} is no longer saturated.
In our computations we neglected the back reaction from the flavor branes (which corresponds to the small number of flavors),
focussed only on trivial embeddings and considered the late-time limit.
Note that the flavor corrections are parametrically small and thus the complexification rate cannot become negative. 

The rest of the paper is organized as follows.
 In Section \ref{review} we review the  proposal of \cite{Brown:2015bva, Brown:2015lvg}; Section \ref{Generalities} covers some generalities of the $D3/Dq$ systems. 
 In section \ref{corrections} we compute corrections to the complexity growth and to the mass of the system. 
We conclude in  Section \ref{Conclusions}.

\section{Review of the Complexity-Action proposal}\label{review}

\noindent A concrete way for computing complexity in QFTs is not yet known. 
However, for some strongly coupled QFTs, such as ${\cal N}=4$ super Yang-Mills, an equivalent gravitational description is available.  
One may then hope that a geometric prescription for evaluating complexity will be easier to define. 
In this article, we will use the proposal of \cite{Brown:2015bva, Brown:2015lvg} .

\begin{figure}[!h]
\centering
\includegraphics[width=75mm]{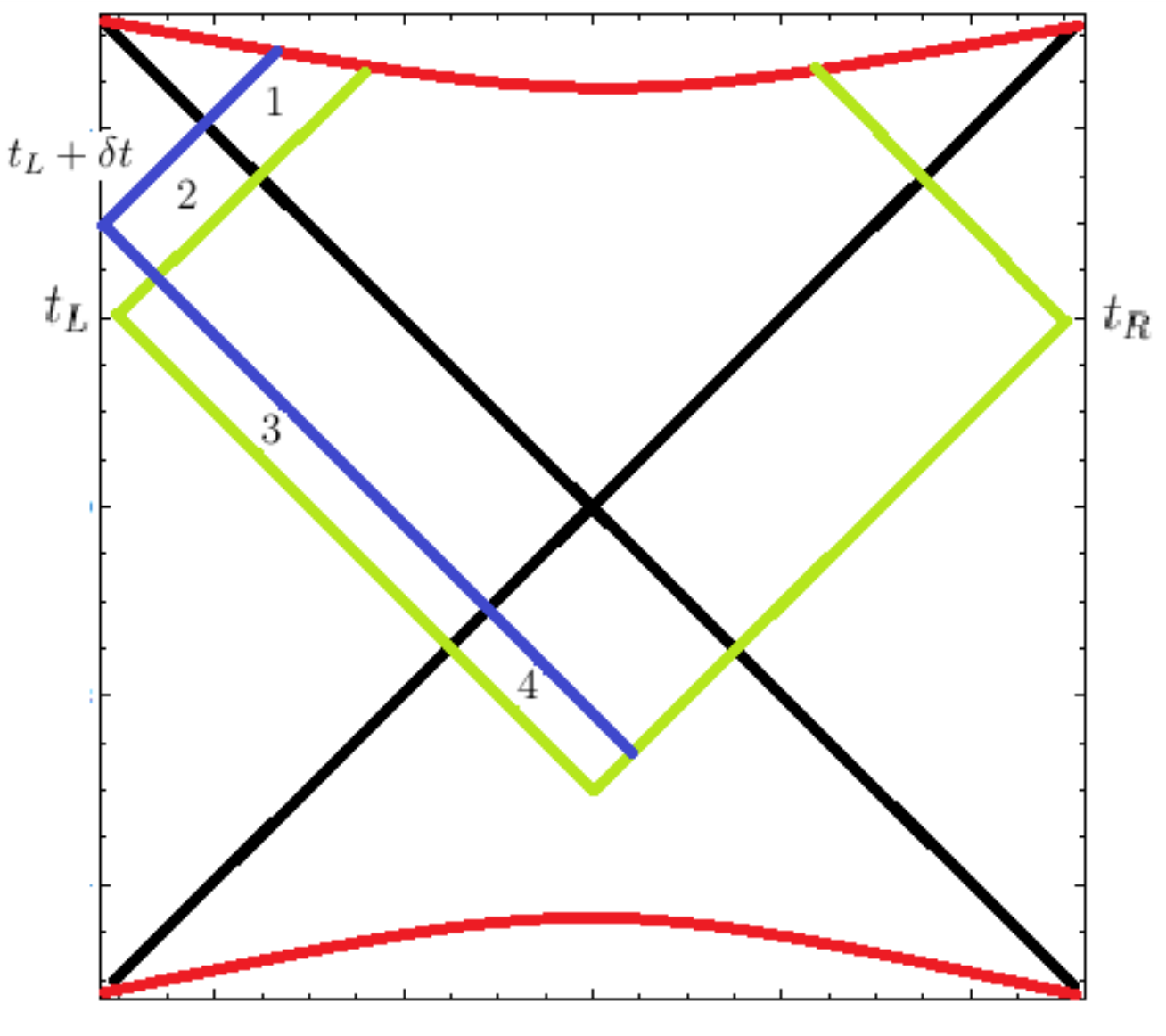}\\
\caption{Penrose diagram of an AdS-Schwarzschild black hole. The red lines represent the future and past singularities, while the black lines crossing the diagram are the event horizons. The area enclosed by the green lines and the future singularity is the WdW patch at times $t_L$ and $t_R$. If $t_L$ is let to evolve infinitesimally the result is the patch shown in blue. This evolution then makes the patch to lose regions 3 and 4 while gaining regions 1 and 2.}
\label{fig:WdWparts}
\end{figure}

The authors of \cite{Brown:2015bva,Brown:2015lvg} provide a prescription for evaluating the complexity of the thermofield double state in the dual gauge theory. For a conformal field theory (CFT) with a holographic dual,
 the finite temperature state is described by the AdS-Schwarzschild spacetime. 
 (We are considering temperatures above the Hawking-Page transition \cite{Hawking:1982dh})
 An important role in the proposal is played by the Wheeler-DeWitt patch, denoted as WdW patch from now on (see Figure \ref{fig:WdWparts}). The proposal states that the complexity $\mathcal{C}$ of the thermofield double-state is given by (\ref{proposalC})
where $S_{WdW}$ is the Einstein-Hilbert action,
\begin{equation}
S=\frac{1}{16\pi G}\int_{\mathcal{M}} \sqrt{-g} \left(\mathcal{R}-2\Lambda\right)+\frac{1}{8\pi G}\int_{\partial\mathcal{M}} \sqrt{h} \mathcal{K},
\end{equation}
evaluated over the WdW patch. As usual, the Einstein-Hilbert action is supplemented by the York-Gibbons-Hawking term (YGH), for the variational problem to be well defined.

This proposal allows one to directly compute $d\mathcal{C}/dt$ and check whether or not the bound \eqref{eq:bound} is respected. Differentiating the holographic complexity is straightforward. Suppose $t_L$ evolves for an infinitesimal amount $\delta t$. Such an evolution changes the WdW patch as shown in Figure \ref{fig:WdWparts}. To compute the change in the action, one  needs to evaluate it on the four regions denoted in  Figure \ref{fig:WdWparts}. However, as already noted in \cite{Brown:2015bva, Brown:2015lvg}, the action evaluated on region 2 is cancelled by that on region 3, while region 4 shrinks to zero in the limit $t_L\gg\beta$.
We will be interested in precisely this limit (large time behavior of the complexity growth).
 So only region 1, the region behind the future singularity, contributes to the rate of change of the holographic complexity. The result presented in \cite{Brown:2015bva, Brown:2015lvg} is the remarkably simple answer
\begin{equation}\label{eq:dCdtSchw}
\frac{d \mathcal{C}}{dt}=\frac{2 \cal{M}}{\pi},
\end{equation}
which exactly saturates the bound \eqref{eq:bound}.\\
\\

\section{D3/Dq systems}\label{Generalities}
\subsection{Generalities}
In this article, we are interested in studying the holographic complexity for a strongly coupled gauge theory with fundamental matter fields (fields transforming under the fundamental representation of the gauge group). 
To this end, we consider D3/Dq systems \cite{Karch:2002sh}. 
These systems are made out of a stack of $N_c$ D3-branes and a number $N_f$ of Dq-branes (the flavour branes).
 Strings stretching between the $N_c$ D3-branes give rise to $\mathcal{N}=4$ SYM, while strings stretching between the D3-branes and the flavour $Dq$-branes introduce fields that transform in the fundamental representation of the gauge group. 
To simplify the discussion, we will focus on the probe limit, where the number of flavor branes is much smaller than that of the color branes: $N_c\gg N_f$.
In this limit, the Dq-branes can be treated as probes, propagating in the spacetime created by the stack of the D3-branes, {\it i.e.}, $AdS_5\times S^5$, without backreaction.

The $Dq$-branes span a $(q+1)$-dimensional worldvolume and thus wrap a $(q+1)$-subspace of $AdS_5\times S^5$.
There are in principle many ways of embedding an $AdS_n\times S^m$ ($n,m \leq 5$) into the background  $AdS_5\times S^5$, {\it i.e.}, several ways of choosing an $S^m$ inside the $S^5$ or an $AdS_n$ inside the $AdS_5$. The embedding is usually specified  by a set of scalar functions determining how the subspaces are chosen inside the 5-sphere and $AdS_5$. For example, for the case of the $D3/D7$ configuration one can consider any of the following embeddings: $AdS_5\times S^3$, $AdS_4\times S^4$ or $AdS_3\times S^5$.\\
However, from all possible embeddings of the form of $AdS_m\times S^n$, only those with $\abs{m-n=2}$ preserve supersymmetry. This means that only the configurations $D3/D3$, $D3/D5$ and $D3/D7$ preserve sypersymmetry and are thus stable. Moreover, all of them can be specified by just one embedding function. 

We will be interested in evaluating the complexity of the thermofield double.
 In the dual gravitational language, this can be achieved by considering $Dq$ branes propagating in the AdS-Schwarzschild spacetime, which describes ${\cal N}=4$ Super Yang Mills at finite temperature. Its metric is given by
\begin{equation}\label{eq:AdSSch}
\begin{aligned}
ds^2=-f(r) dt^2+\frac{dr^2}{f(r)}+r^2 d\Omega_3^2+L^2d\Omega_5^2;\\
f(r)=1+\frac{r^2}{L^2}-\frac{M}{r^2} \,,\
\end{aligned}
\end{equation}
where $M=\frac{8 G}{3\pi}\mathcal{M}$. Apart from the dependence on $L$, the radius of curvature of both the $AdS_5$ and the $S^5$ spaces, the AdS-Schwarzschild metric also depends on an additional parameter $M$ which is proportional to the mass of the black hole. The Penrose diagram of the AdS-Schwarzschild spacetime is depicted in Fig.1.\\

To evaluate the contribution to the complexity of the state from the flavor degrees of freedom in the large $N_c$ and  large't Hooft coupling $\lambda$ limit, we simply need to evaluate the action for the propagation of the probe Dq branes in the AdS-Schwartzchild background on the WdW patch. The action which governs the propagation of the Dq branes is the DBI action:\footnote{The Euclidean DBI action has a positive sign. Also, we will denote the Euclidean action as $I$ instead of $S$ to avoid confusion with entropy.
Note that the variation of this action is proportional to just $\delta g_{\mu\nu}$, so no boundary terms are needed here to make the variational problem well defined.}
\begin{equation}\label{eq:DBIaction}
S_{DBI}=-N_f T_{Dq}\int \sqrt{-g_{Dq}},\,
\end{equation}
where the tension of the $Dq$-brane is given by
\begin{equation}
T_{Dq}=\frac{1}{(2\pi \mathit{l}_s)^q g_s \mathit{l}_s}.
\end{equation}
The string length $\ell_s$ and the string coupling constant $g_s$ are related to the `t Hooft coupling $\lambda$ and the colour degrees of freedom $N_c$ through
\begin{equation}
\lambda=g_{YM}^2 N_c=2\pi g_s N_c, \quad \quad L^4=4\pi g_s N_c \mathit{l}_s^4\, \, ,
\end{equation}
where $L$ denotes the $AdS$ radius of curvature as above. In \eqref{eq:DBIaction} $g_{Dq}$ denotes the determinant of the induced metric of the $Dq$ branes, which depends on the details of the embedding.

The embeddings we consider in this article, are the trivial embeddings, and correspond to adding massless flavor matter in the $N=4$ SYM Lagrangian. As explained above, the asymptotic form of the induced metric will be $AdS_m\times S^n$. Evaluating the DBI action on asymptotically AdS geometries leads to divergences which can be  treated with holographic renormalization \cite{Henningson:1998gx, Henningson:1998ey}. Holographic renormalization for the case of D3/Dq systems was studied in \cite{Karch:2005ms}. For technical reasons it is convenient to express the AdS-Schwarzschild metric in Fefferman-Graham coordinates
\begin{equation}\label{eq:FFGGeneralMetric}
ds^2=L^2\Bigg\{\frac{dz^2}{z^2}+\frac{L^2}{4z^2} \left[1-\frac{z^4}{L^4}\left(1+4\frac{M}{L^2}\right)\right]^2 \frac{d\tau^2}{F(z,M)}+\frac{F(z,M)}{4 z^2}d\Omega_3^2+d\Omega_5^2\Bigg\}\,,
\end{equation}
where 
\begin{equation}
F(z,M)=L^2-2z^2+\frac{z^4}{L^2}\left(1+4\frac{M}{L^2}\right).
\end{equation}
The boundary of AdS is now at $z=0$, while the horizon is mapped to 
\begin{equation}\label{eq:zHDef}
z_H\equiv z(r=r_H)=\frac{L^2}{\sqrt{L^2+2r_H^2}}.
\end{equation}
The radial coordinates $(z,r)$ are related to one another as follows:
\begin{equation}
z(r)=\frac{L^2}{\left[L^2+2r^2+2\sqrt{r^4+L^2r^2-L^2M}\right]^{1/2}}, \qquad r^2=L^2\frac{F(z,M)}{4 z^2}\,.
\end{equation}

The trivial embeddings considered in this paper are described by induced metrics with asymptotics of the form $AdS_m\times S^n$, where $m+n=q+1$ and
\begin{equation}\label{eq: GenFGMetric}
ds_{Dq}^2=L^2\Bigg\{\frac{dz^2}{z^2}+\frac{L^2}{4z^2} \left[1-\frac{z^4}{L^4}\left(1+4\frac{M}{L^2}\right)\right]^2 \frac{d\tau^2}{F(z,M)}+\frac{F(z,M)}{4 z^2}d\Omega_{n-2}^2+d\Omega_{q-n+1}^2\Bigg\}.
\end{equation}
As explained above, we will use Holographic Renormalization in order to deal with the divergent contributions in $\int \sqrt{g_{Dq}}$. The procedure consists of the following steps: firstly, we introduce a cutoff surface at $z=\epsilon$ and define covariant counterterms on the $z=\epsilon$ surface such that the divergences are cancelled. Then, we take the limit $\epsilon\rightarrow 0$ to remove the cutoff. The appropriate counterterms were worked out in \cite{Karch:2005ms} and are of two classes; the ones needed to regulate the volume part of the integral and the ones required to regulate the contributions from the embedding functions. For trivial embeddings only the former type of counterterms appear since the embedding functions are zero. As a result, for the induced metrics quoted in \eqref{eq: GenFGMetric} the following counterterms are required:
\begin{equation}\label{eq:GeneralCounter}
\begin{aligned}
I^{ren}=I_{DBI}+I_{count};\quad\quad I_{count}=N_f T_{Dq}\int \sqrt{\gamma}(L_1+L_2)=N_f T_{Dq}\int \sqrt{\gamma}(-a+b \mathcal{R}_\gamma)\\
a= \begin{cases} 
      L/4 & \text{for } AdS_5 \\
      L/3 & \text{for }AdS_4 \\
      L/2 & \text{for }AdS_3 \\
      L   & \text{for }AdS_2 
   \end{cases}
   \quad \quad \quad b= \begin{cases} 
      L^3/48 & \text{for } AdS_5 \\
      L^3/12 & \text{for }AdS_4 \\
      0 & \text{for }AdS_3 \\
      0   & \text{for }AdS_2 
   \end{cases}
\end{aligned}
\end{equation}
where $\mathcal{R}_\gamma$ is the Ricci scalar associated with the induced metric $\gamma$ on the constant $z$ surface.

\section{Complexity and energy of the D3/Dq systems}\label{corrections}

In this section we address the main question of this article. We compute the time derivative of the DBI action over the WdW patch and express it in terms of the energy of the system, in order to check if \eqref{eq:bound} is respected. 
We first study in detail the D3/D7 and D3/D5 configurations and then discuss the general case. For each system, we start by working out the correction to the energy due to the flavor branes and then compute the rate of change of the complexity.


\subsection{Complexity and energy of the D3/D7 system}\label{D3D7}

\subsubsection{Energy of the system}

The thermodynamic properties of a system are derived from its Euclidean action, which in this case in the DBI action, $I_{D7}$. The correction to the free energy of the black hole is given by $\delta F=T I_{D7}$ and the energy is obtained from the thermodynamic relation
\begin{equation}
\delta \mathcal{M}=\delta F+T \delta S, \quad \quad \quad \delta S=\frac{\partial \delta F}{\partial T}.
\end{equation}
In terms of inverse temperature $\beta=1/T$, the above relation can be expressed as
\begin{equation}\label{eq:EnergyFormula}
\delta \mathcal{M}=\delta F+\beta \frac{\partial \delta F}{\partial \beta}.
\end{equation}
To compute $\delta \mathcal{M}$ we thus need to evaluate the Euclidean DBI action on the $D7$-brane configuration:
\begin{equation}
\begin{aligned}
I_{D7}=N_f T_{D7}\frac{L^9}{16}\int_0^\beta d\tau\int d\Omega_3\int d\Omega_3\int_0^{z_H}\frac{F(z)}{z^5}\left[1-\frac{z^4}{L^2}\left(1+4\frac{M}{L^2}\right)\right]\\
=N_f T_{D7} \frac{L^9}{16} V_{\Omega_3}^2 \beta \left[\frac{-L^2}{4 z^4}+\frac{1}{z^2}+\frac{(L^2+4M)z^2}{L^6}-\frac{(L^2+4M)^2 z^4}{4L^{10}}\right]_0^{z_H}.
\end{aligned}
\end{equation}
As anticipated above, the action diverges when $z\rightarrow 0$.
Introducing a cutoff at $z=\epsilon$ and evaluating the relevant counterterms from \eqref{eq:GeneralCounter} yields
\begin{equation}
I_{count}=N_f T_{D7} V_{\Omega_3}^2 \beta \left[-\frac{L^{11}}{64 \epsilon^4}+\frac{L^9}{16\epsilon^2}+\mathcal{O}(\epsilon^2)\right],
\end{equation}
which exactly cancels the divergences of $I_{D7}$ without introducing any finite contribution. The final result is 
\begin{equation}
I^{ren}_{D7}=N_f T_{D7} \frac{L^9}{16} V_{\Omega_3}^2 \beta \left[\frac{-L^2}{4 z_H^4}+\frac{1}{z_H^2}+\frac{(L^2+4M)z_H^2}{L^6}-\frac{(L^2+4M)^2 z_H^4}{4L^{10}}\right].
\end{equation}

To compute the thermodynamic quantities we're interested in, we need to write $I^{ren}_{D7}$ as a function of $\beta$. To do so we use \eqref{eq:zHDef} to relate $z_H$ with $r_H$, where $r_H$ is the position of the horizon of the AdS-Schwartzchild black hole in the original coordinates \eqref{eq:AdSSch} and is related to the temperature as \cite{Witten:1998zw},
\begin{equation}\label{eq:rH(beta)}
r_H(\beta)=\frac{L^2\pi+\sqrt{L^4\pi^2-2L^2\beta^2}}{2\beta}=L \frac{x+\sqrt{x^2-2}}{2}.
\end{equation}
Note that there is a minimum temperature allowed, namely $T=\frac{\sqrt{2}}{\pi L}$. This is the temperature below which black holes cannot exist.\\
Solving $f(r_H)=0$, one finds that
\begin{equation}
r_H^2=L^2 \frac{-1+\sqrt{1+4M/L^2}}{2},
\end{equation}
which, together with \eqref{eq:zHDef}, leads to
\begin{equation}\label{eq:zHInrH}
z_H=\frac{L^2}{(L^2+2r_H^2)^{1/2}} \quad \rightarrow \quad z_H=\frac{L}{\left(1+4\frac{M}{L^2}\right)^{1/4}}.
\end{equation}
Substituting into our result for $I_{D7}^{ren}(z_H,M)$ results in
\begin{equation}
I^{ren}_{D7}=\frac{N_f T_{D7} L^7 V_{\Omega_3}^2 \beta}{32}\left[4\left(1+2\frac{r_H^2}{L^2}\right)-\left(1+2\frac{r_H^2}{L^2}\right)^2\right].
\end{equation}
It is easy to express $I_{D7}^{ren}(\beta)$ in terms of the inverse temperature $\beta$ by using \eqref{eq:rH(beta)}. Applying \eqref{eq:EnergyFormula} then leads to the following expression for the energy of the $D7$ system
\begin{equation}\label{eq:ED7}
\begin{aligned}
\delta \mathcal{M}_{D7}=\frac{N_f T_{D7} L^7 V_{\Omega_3}^2}{32}H_{D7}(\beta).\\
H_{D7}(\beta)\equiv 6\left[\frac{L^4\pi^4}{\beta^4}-\frac{L^2\pi^2}{\beta^2}+\frac{L^2\pi^3\sqrt{L^4\pi^2-2L^2\beta^2}}{\beta^4}\right].
\end{aligned}
\end{equation}
In the planar limit, $L/\beta\rightarrow\infty$, this agrees with eq. (4.28) in \cite{Mateos:2007vn} (see also \cite{Karch:2006bv} for a similar computation for massive enbeddings).

\subsubsection{Complexity}
\begin{figure}[!h]
\centering
\includegraphics[width=75mm]{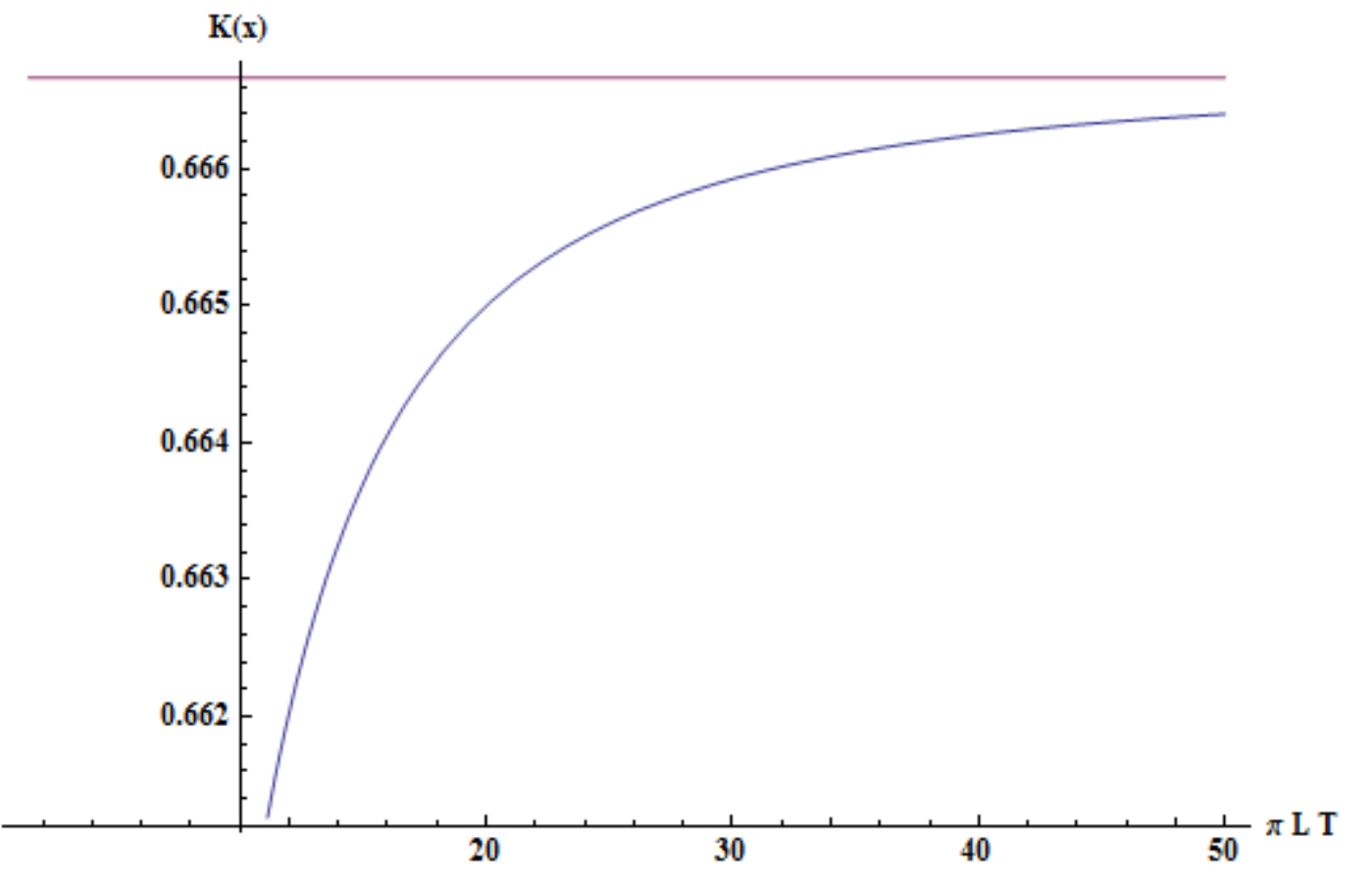}\\
\caption{Plot of the function $K_{D7}(x)$ starting from the minimum value $x_{min}=\sqrt{2}$. The horizontal orange line is the value to which it asymptotes, namely $2/3$.}
\label{fig:KD7}
\end{figure}

Here we discuss the complexity computation. The Penrose diagram of the D3/D7 system is still the one shown in Figure \ref{fig:WdWparts}, so our integral will split into the same 4 regions. The difference is that now our action is\footnote{Recall that the Lorentzian action has negative sign.}
\begin{equation}
\delta\mathcal{C}=S_{DBI}=-N_f T_{D7}\int_{WdW} \sqrt{-g}.
\end{equation}
Note that no surface terms are needed since the variation $\delta S_{DBI}$ contains no terms depending on $\delta (\partial_\sigma g_{\mu\nu})\Big|_{\partial\mathcal{M}}$ . With our action, the integrals from parts 2 and 3 again cancel each other out, and the region 4 doesn't contribute either because it shrinks to zero size\footnote{In the Einstein gravity case studied in \cite{Brown:2015lvg} a topological argument is needed to rule this part out because the integrand there is $\mathcal{R}$; since our integral is just a volume for us this argument is trivial.}. So we are only left with region 1, which is bounded by the surfaces $r=0$ and $r=r_H$. Working with the metric as in \eqref{eq:AdSSch}, the integrand is
\begin{equation}
\sqrt{-g}=r^3 L^3 .
\end{equation}
The time derivative of the action is then simply
\begin{equation}\label{eq:dSdt}
\begin{aligned}
\frac{d S_{DBI}}{dt}=-N_f T_{D7}\frac{d}{dt}\int\sqrt{-g}=-N_f T_{D7} L^3 \int dr r^3 \int d\Omega_3 \int d\Omega_3\\
=-N_f T_{D7} L^3 V_{\Omega_3}^2 \frac{r_H^4}{4}=-N_f T_{D7} L^7 V_{\Omega_3}^2 \frac{r_H^4}{4 L^4}.
\end{aligned}
\end{equation}
We would like to express our result for the complexity as a function of the temperature and the energy of the system. To introduce the energy into the last equation we use \eqref{eq:ED7} to write the overall factor in \eqref{eq:dSdt} as
\begin{equation}\label{eq:FactorInD7}
N_f T_{D7} L^7 V_{\Omega_3}^2=\frac{32 \,\, \delta M_{D7}}{H_{D7}(\beta)}.
\end{equation}
So, using \eqref{eq:FactorInD7} and \eqref{eq:rH(beta)} yields
\begin{equation}\label{eq: D7Complex}
\begin{aligned}
\frac{d(\delta \mathcal{C})}{dt}=\frac{d S_{DBI}}{dt}=-\frac{\delta \mathcal{M}}{\pi}K_{D7}(x),\\
K_{D7}(x)\equiv \frac{8 r_H(\beta)^4}{H_{D7}(\beta)}=\frac{1}{12}\frac{x^2\left[1+\sqrt{1-\frac{2}{x^2}}\right]^4}{x^2\left[1+\sqrt{1-\frac{2}{x^2}}\right]-1}, \quad \quad \quad x=\pi L T.
\end{aligned}
\end{equation}
Note that there is a minimum value $x$ can take, being $x_{min}=\sqrt{2}$. The function $K(x)$ is plotted on Figure \ref{fig:KD7}. The function is monotonically increasing, positive and ranging between the value $1/6$ at the minimum and asymptotically approaching $2/3$.

\noindent Due to the minus sign present in \eqref{eq: D7Complex} the correction lowers the speed at which the system complexifies, so the bound is respected but not saturated.

\subsection{Complexity and energy of the D3/D5 system}\label{D3D5}

\subsubsection{Energy of the system}

To compute the correction to the energy of the D3/D5 due to the flavor D5 branes in the probe limit, we will follow exactly the same steps as in section 4.1. The Euclidean action is in this case given by
\begin{equation}
\begin{aligned}
I_{D5}=T_{D5} N_f \int \sqrt{g}=T_{D5} N_f \beta V_{\Omega_2}^2\frac{L^7}{8}\int_{0}^{z_H} dz\frac{\left[1-\frac{z^4}{L^4}\left(1+4\frac{M}{L^2}\right)\right]}{z^4}\sqrt{F(z)}\\
=-\frac{T_{D5} N_f \beta V_{\Omega_2}^2 L^7}{8}\left[\frac{F(z)^{3/2}}{3L^2 z^3}\right]_\epsilon^{z_H}\,,
\end{aligned}
\end{equation}
with divergenent terms of the form
\begin{equation}
I_{D5}^{div}= -T_{D5} N_f \beta V_{\Omega_2}^2 \left[\frac{L^8}{24 \epsilon^3}-\frac{L^6}{8\epsilon^2}+\mathcal{O}(\epsilon)\right].
\end{equation}
The relevant counterterms are
\begin{equation}
\begin{aligned}
I_{count}=N_f T_{D5}\int \sqrt{\gamma} (L_1+L_2)\quad \quad \longrightarrow \quad \quad I^{ren}=I+I_{count}\\
L_1=\frac{-L}{3}, \quad \quad \quad L_2=\frac{L^3}{12}\mathcal{R}_\gamma.
\end{aligned}
\end{equation}
Just as in the D3/D7 case, the holographic renormalization procedure removes the divergent parts without adding any finite terms. The final result is:
\begin{equation}
I_{D5}^{ren}=-\frac{T_{D5} N_f \beta V_{\Omega_2}^2 L^5}{8}\left[\frac{F(z_H)^{3/2}}{3 z_H^3}\right].
\end{equation}
Using \eqref{eq:zHInrH} it's immediate to see that
\begin{equation}\label{eq:FAtzH}
F(z_H)=L^2 \frac{4 r_H^2/L^2}{1+2 r_H^2/L^2}, \quad \quad z_H^3=\frac{L^3}{(1+2 r_H^2/L^2)^{3/2}},
\end{equation}
which allows us to write the renormalized action as
\begin{equation}
I_{D5}^{ren}=-\frac{T_{D5} N_f \beta V_{\Omega_2}^2 L^5}{3} \frac{r_H^3}{L^3}.
\end{equation}
The correction to the free energy of the D3/D5 system is
\begin{equation}
\delta F_{D5}=-\frac{T_{D5} N_f V_{\Omega_2}^2 L^5}{3} \frac{r_H^3}{L^3}.
\end{equation}
With the help of \eqref{eq:rH(beta)} we obtain the free energy as a function of the inverse temperature, $F_{D5}(\beta)$ and use the standard thermodynamic relations \eqref{eq:EnergyFormula} to obtain
\begin{equation}\label{eq:ED5}
\begin{aligned}
\delta \mathcal{M}=\frac{T_{D5} N_f V_{\Omega_2}^2 L^5}{3}\: H_{D5}\left(x\right),\\
H_{D5}\left(x\right)=\frac{2 x^4+2x^3\sqrt{x^2-2}-2x^2-1}{2\sqrt{x^2-2}},&\quad \quad x\equiv \pi L T.
\end{aligned}
\end{equation}

\subsubsection{Complexity}

Let's now see how the complexity is related to the energy in the D3/D5 system. The arguments made in section 4.1.2 regarding the contribution of the different parts of the WdW patch are still valid, and clearly the first equality in \eqref{eq:dSdt} is still true (changing $T_{D7}\leftrightarrow T_{D5}$), the only difference being the explicit form of $\sqrt{-g}$. The induced metric is in this case asymptotically $AdS_4\times S^2$: 
\begin{equation}
ds^2=-f(r)dt^2+\frac{dr^2}{f(r)}+r^2 d\Omega_2^2+L^2 d\Omega_2^2,
\end{equation}
with the determinant
\begin{equation}
\sqrt{-g}=r^2 L^2 .
\end{equation}
Following exactly the same steps which lead us to \eqref{eq:dSdt} and dividing by $V_x$ to obtain a density, leads to
\begin{equation}
\frac{d S_{DBI}}{dt}=-\frac{N_f T_{D5}V_{\Omega_2}^2 L^5}{3} \frac{r_H^3}{L^3}.
\end{equation}
Similar to the D3/D7 case, the factor multiplying $r_H^3/L^3$ in the equation above can be expressed in terms of the energy of the system
\begin{equation}
\frac{T_{D5} N_f V_{\Omega_2}^2 L^5}{3}=\frac{\delta M}{H_{D5}(x)}.
\end{equation}
This together with \eqref{eq:rH(beta)} transforms the equation for $dS_{DBI}/dt$ into
\begin{equation}
\begin{aligned}
\frac{d \,\delta\mathcal{C}}{dt}&=-\frac{\delta \mathcal{M}}{\pi}K_{D5}(x),\\
K_{D5}(x)=\left(\frac{x+\sqrt{x^2-2}}{2}\right)^3\,H_{D5}^{-1}(x), &\quad \quad \text{with }x=\pi L T.
\end{aligned}
\end{equation}
Note again, that there is a minimum value allowed for $x$, namely $x=\sqrt{2}$. The function is positive, monotonically increasing and ranges between $0$ at the minimum and the asymptotic value $1/2$.\\
\begin{figure}[!h]
\centering
\includegraphics[width=75mm]{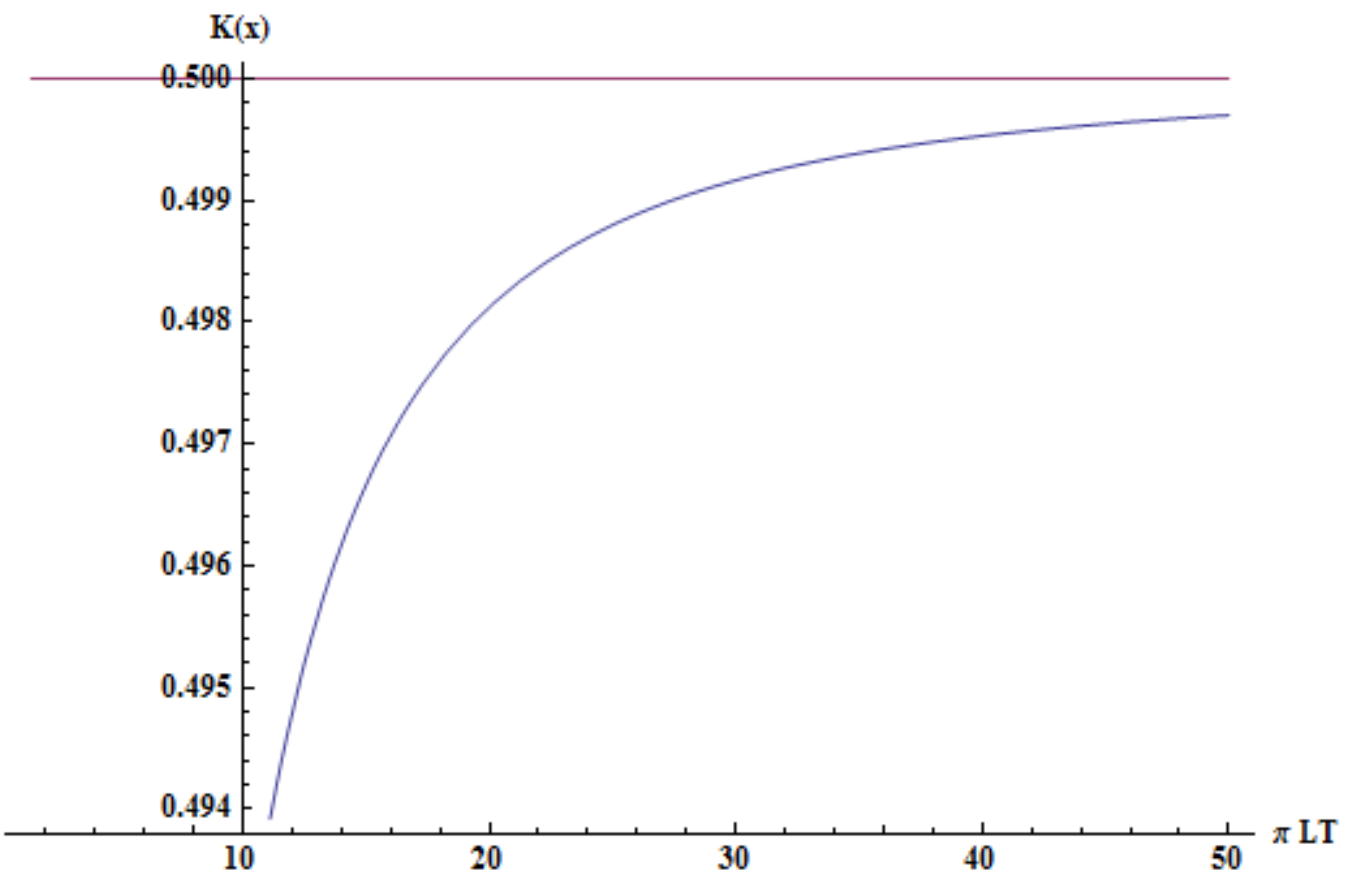}\\
\caption{The function $K_{D5}$ vs temperature, starting from the minimum value $x_{min}=\sqrt{2}$. The horizontal orange line is the value to which it asymptotes, namely $1/2$.}
\label{fig:H}
\end{figure}

\subsection{The general case: D3/Dq systems}\label{D3Dq}

Having gained some insight from the detailed study of the $D3/D5$ and the $D3/D7$ systems, we move on to consider the generic $D3/Dq$ system. As we will see, the qualitative features of the complexity of the thermofield double state in the presence of flavour matter fields, remain the same for both stable and unstable (non-supersymmetric) configurations.

\subsubsection{The energy of the D3/Dq systems.}

As discussed above, the different embeddings of the $Dq$-branes are submanifolds of the $AdS_5\times S^5$ generated by the background $D3$-branes, with the asymptotic form of $AdS_n\times S^m$ where $m+n=q+1$. Regarding the energy computation, all the divergent parts in the Euclidean action come from the $AdS_n$ part of the manifold.
The induced metric on the $Dq$ branes is given in \eqref{eq: GenFGMetric} and its determinant is equal to:
\begin{equation}\label{eq: GenFGMeasure}
\sqrt{g}=\frac{L^{q+2}}{2^{n-1}} z^{-n}\left[1-\frac{z^4}{L^4}\left(1+4\frac{M}{L^2}\right)\right]F(z)^{\frac{n-3}{2}} .
\end{equation}

It is straightforward to evaluate the Euclidean DBI action $I_{Dq}$ to obtain
\begin{equation}\label{eq:IDqPre}
I_{Dq}=N_f T_{Dq}\int \sqrt{g}=N_f T_{Dq} \Big[-\frac{L^q}{2^{n-1}(n-1)}\frac{ F(z)^{\frac{n-1}{2}}}{z^{n-1}}\Big]_0^{z_H}\beta V_{\Omega_{n-2}}V_{\Omega_{q-n+1}}
\end{equation}
To proceed it will be convenient to separately analyze the cases where the $AdS_n$ part of the embedding is of even or odd dimensionality.

\paragraph{When $n$ is an even integer.}

As one can see from \eqref{eq:IDqPre}, for $n$ even, the Euclidean action is given in terms of the metric function $F(z)$ elevated to a half-integer power. The behaviour of $I_{Dq}$ for small $z$ can be split into two types of contributions
\begin{equation}
I_{Dq}^{even}\Big|_{z\rightarrow 0}\rightarrow f\left(z\right)+g\left(\frac{1}{z}\right),
\end{equation}
where $f(z)$ and $g({1\over z})$ represent polynomial functions in $z$ and ${1\over z}$ respectively, with vanishing zeroth order terms. $f(z)$ then vanishes when evaluated at $z\rightarrow 0$, while $g({1\over z})$ is divergent but its divergences are exactly cancelled by the relevant counterterms and no constant piece is introduced. The result is then given by contributions from just the horizon as
\begin{equation}
\begin{aligned}
I_{Dq}=-N_f T_{Dq}\frac{L^q}{2^{n-1}(n-1)}\frac{ F(z_H)^{\frac{n-1}{2}}}{z_H^{n-1}}\beta V_{\Omega_{n-2}}V_{\Omega_{q-n+1}}\\
=-N_f T_{Dq}\frac{L^q}{n-1} \left(\frac{r_H}{L}\right)^{n-1}\beta V_{\Omega_{n-2}}V_{\Omega_{q-n+1}},
\end{aligned}
\end{equation}
where we have used \eqref{eq:FAtzH} in the last equality. \\
We can now write the free energy $F_{Dq}=T I_{Dq}$ and use \eqref{eq:EnergyFormula} to obtain the energy of the system as,
\begin{equation}\label{eq:EnerDq}
\begin{aligned}
\delta \mathcal{M}=N_f T_{Dq} \frac{L^q}{(n-1)}H_{Dq}(x) V_{\Omega_{n-2}}V_{\Omega_{q-n+1}},\\
H_{Dq}(x)=\left(\frac{r_H(x)}{L}\right)^{n-2}\frac{2+(n-2)x\sqrt{x^2-2}+(n-2)x^2}{2\sqrt{x^2-2}}, \quad \quad \frac{r_H(x)}{L}=\frac{x+\sqrt{x^2-2}}{2}.
\end{aligned}
\end{equation}

\paragraph{When $n$ is an odd integer.}
In this case $F(z)$ is elevated to an integer power, and the result is a polynomial in even powers of $z$, {\it i.e.},
\begin{equation}
F(z)^{\frac{n-1}{2}}=A_0+A_2 z^2+\cdots + A_{2(n-1)} z^{2(n-1)}\,,
\end{equation}
This implies that the quantity $F^{\frac{n-1}{2}}/z^{n-1}$ in \eqref{eq:IDqPre} contains a constant term, independent from $z$. Once more the divergent terms at the boundary $z=0$ are precisely cancelled by the relevant counterterms and the Euclidean action is given by
\begin{equation}
I_{Dq}=-N_f T_{Dq}\frac{L^q}{2^{n-1}(n-1)}\left[\frac{ F(z_H)^{\frac{n-1}{2}}}{z_H^{n-1}}-c_0\right]\beta V_{\Omega_{n-2}}V_{\Omega_{q-n+1}}.
\end{equation}
Clearly the constant term, indicated by $c_0$, is cancelled by the same $z$-independent term in $\frac{ F(z_H)^{\frac{n-1}{2}}}{z_H^{n-1}}$.

In practice, there exist only two non-trivial embeddings in this class: those which asymptote to $AdS_3$ and those which asymptote to $AdS_5$. The latter case was addressed in the context of the $D3/D7$ system, we only need to consider the $AdS_3$ case. From \eqref{eq: GenFGMetric} and \eqref{eq: GenFGMeasure} we can see that we are now working with
\begin{equation}
\begin{aligned}
ds^2_{Dq}=L^2\Bigg\{\frac{dz^2}{z^2}+\frac{L^2}{4z^2} \left[1-\frac{z^4}{L^4}\left(1+4\frac{M}{L^2}\right)\right]^2 \frac{d\tau^2}{F(z,M)}+\frac{F(z,M)}{4 z^2}d\theta^2+d\Omega_m^2\Bigg\},\\
\sqrt{g_{Dq}}=\frac{L^4}{4} z^{-3}\left[1-\frac{z^4}{L^4}\left(1+4\frac{M}{L^2}\right)\right] dz\; d\tau\; d\theta (L^m d\Omega_m).
\end{aligned}
\end{equation}
It is straightforward to apply the general result above to the case $n=3$ to obtain:
\begin{equation}
I_{Dq}=-\frac{N_f T_{Dq} L^q}{4} \,  \beta V_{\Omega_1} V_{\Omega_{q-2}} \,\left(1+\frac{2 r_H^2}{L^2}\right)\,,
\end{equation}
where we used the relation between $(z_H, F(z_H))$ and $r_H$ from \eqref{eq:zHDef}.
Evaluating \eqref{eq:EnergyFormula} then yields
\begin{equation}
\begin{aligned}
\delta \mathcal{M}=\frac{N_f T_{Dq}L^2 V_{\Omega_1}}{2} H_{Dq}(x),\\
H_{Dq}(x)=\frac{x^3+x^2\sqrt{x^2-2}}{\sqrt{x^2-2}}.
\end{aligned}
\end{equation}

\subsubsection{Complexity of the $D3/Dq$ system.}

\paragraph{When $n$ is an even integer.}
We follow exactly the same steps as in the previous sections to evaluate the time derivative of the DBI action $S_{DBI}=-N_f T_{Dq}\int\sqrt{-g}$, which is given by
\begin{equation}\label{eq:ComplexGenQ}
\frac{d S_{DBI}}{dt}=-\frac{N_f T_{Dq} L^q}{n-1} \left(\frac{r_H}{L}\right)^{n-1}V_{\Omega_{n-2}}V_{\Omega_{q-n+1}}.
\end{equation}
As usual, we can solve \eqref{eq:EnerDq} for $N_f T_{Dq}L^q$ to write this derivative as
\begin{equation}
\begin{aligned}
\frac{d\,\delta\mathcal{C}}{dt}=-\frac{\delta \mathcal{M}}{\pi H_{Dq}(x)} \left(\frac{r_H}{L}\right)^{n-1}=-\frac{\sqrt{x^2-2}\left(x+\sqrt{x^2-2}\right)}{2-(n-2)x\sqrt{x^2-2}+(n-2)x^2}\, \frac{\delta \mathcal{M}}{\pi}\equiv -K_{Dq}(x) \frac{\delta \mathcal{M}}{\pi}
\end{aligned}
\end{equation}

\paragraph{When $n$ is an odd integer.}

For odd $n$ we only need to consider $n=3$ and focus on embeddings which asymptote to $AdS_3$ as in 4.3.1. Similarly to the previous sections we obtain
\begin{equation}
\begin{aligned}
\frac{d\,\delta \mathcal{C}}{dt}=\frac{d S_{DBI}}{\pi dt}=-\frac{N_f T_{Dq} L^q}{2\pi} \left({r_H\over L}\right)^2 V_{\Omega_1}\,V_{\Omega_{q-2}},
\end{aligned}
\end{equation}
which coincides with equation \eqref{eq:ComplexGenQ} for $n=3$. As usual, we can solve the energy equation to express the numerator as a function of $\delta M$. This produces the final result
\begin{equation}
\frac{d\,\delta\mathcal{C}}{dt}=-\frac{r_H^2}{L^2} \frac{\delta \mathcal{M}}{\pi H_{Dq}(x)}=-\frac{\sqrt{x^2-2}(x+\sqrt{x^2-2})}{4x^2} \frac{\delta \mathcal{M}}{\pi}
\end{equation}

Clearly, the correction to the complexity due to the probe, flavor branes is negative and monotonically decreasing for all the D3/Dq systems.

\section{Conclusions}\label{Conclusions}
Introducing fundamental matter leads to a correction term to the left-hand side of \eqref{eq:dCdtSchw}
which is negative.
It is interesting that the growth of quantum complexity in systems with fundamental matter seems to be slower than 
that with just adjoint matter.
It would be interesting to compare this with a direct computation in field theory.
Note that the presence of extra matter in the bulk was shown to reduce the rate of complexity growth in \cite{Brown:2015lvg}.

It would be interesting to compute the flavor corrections to the complexification rate
using the complexity-volume proposal \cite{Stanford:2014jda}.
It is not immediately clear to us how to generalize this proposal to include flavor corrections.

It would also be interesting to study the behavior of the quantum complexity growth in non-conformal field theories. In gravity, one could investigate asymptotically AdS domain wall solutions or general Dp/Dq systems.

\section*{Acknowledgements}
M.K. acknowledges support from a Marie-Curie fellowship with project no 203972 within the European Research and Innovation Programme EU H2020/2014-2020.


\end{document}